\documentclass[12pt]{article}

\usepackage{graphics}
\usepackage{amsmath}
\def\ni{\noindent}
\def\ph{{\phantom{...}}}
\def\={\phantom{..} = \phantom{..}}

\def\+{\phantom{..} + \phantom{..}}
\def\>{\phantom{..} > \phantom{..}}
\def\<{\phantom{..} < \phantom{..}}
\def\-{\phantom{..} - \phantom{..}}

\def\dotskip{\vskip0.2in \centerline{ . . . } \vskip0.2in}

\def\bq{\begin{quote}}
\def\eq{\end{quote}}
\def\be{\begin{equation}}
\def\ee{\end{equation}}
\def\bar{\begin{eqnarray}}
\def\ear{\end{eqnarray}}
\def\no{\nonumber}

\def\ea{{\em et. al}}
\def\rv{random variable}
\def\rv{random variables}
\def\Sch{Schr{\"o}dinger}
\def\Schism{Schr{\"o}dingerism}

\def\Schists{Schr{\"o}dingerists}
\def\partist{particle-ist}
\def\partism{particle-ism}

\def\Schs{Schr{\"o}dinger's}
\def\Schseqn{Schr{\"o}dinger's equation}

\def\paism{particle-ism}

\def\vN{von Neumann}

\def\wf{wavefunction}

\def\Re{\hbox{Re}}
\def\rv{random variable}
\def\rvs{random variables}
\def\nai{na{\"i}ve}

\def\hpsi{\hat{\psi}}
\def\Re{\hbox{Re}}

\def\Plb{P\left[\,}
\def\Prb{\,\right]}

\title{\bf On Non-Linear Quantum Mechanics\\[1in] and the Measurement Problem\\[1in] II. The Random Part of the Wavefunction\\[2in]}

\author{W. David Wick\footnote{email: wdavid.wick@gmail.com}}

\begin{document}
\maketitle
\pagebreak

\section*{Abstract}

In the first paper of this series, I introduced a non-linear, Hamiltonian, generalization of \Sch's theory that blocks formation of macroscopic dispersion (``cats").
But that theory was entirely deterministic, and so the origin of random outcomes in experiments 
such as Stern-Gerlach or EPRB was left open. Here I propose that \Sch's\ \wf\ has a random component and demonstrate
that such an improvised stochastic theory can violate Bell's inequality. 
Repeated measurements and the back-reaction on the microsystem are discussed in a toy example. 
Experiments that might falsify
the theory are described. 
\vfill\pagebreak

\section{Introduction}
\subsection{Bell's Inequality\label{Bell_sect}}

In 1964, John Bell published a short paper, \cite{bell},
making the claim: QM is incompatible with ``hidden variables." He drew this conclusion from a short argument about
the predictions of QM for the Einstein-Podolsky-Rosen thought-experiment in David Bohm's concrete version, presented near the end of his 1951 textbook, \cite{bohm51}.  
I'll reprise the experiment, now abbreviated EPRB, 
and sketch Bell's argument and later refinements in this section.

EPRB is a dual version of Stern-Gerlach in which a ``pair of particles" initially in a ``total spin-zero" symmetrical state is sent to two SG apparatus, 
say at equal distances to right and left of the $x$-axis.
Since Bell used `A' and `B' for the variables measured by the apparatus, they have become known as operating in Alice's and Bob's laboratories. Alice and Bob choose the orientation of 
the external magnetic field in both labs in the $y$-$z$ plane in directions defined by unit vectors written `a' and `b'. 
Now let us assume, as is standard, that this set-up produces a joint \vN\ measurement of
the two spin operators $a\cdot\sigma_1$ and $b\cdot\sigma_2$ ($\sigma_1$ being the triple of 2x2 Pauli spin operators representing ``the spin of particle one" and ditto for ``particle two").
An easy calculation yields

\be
<\psi|\,a\cdot\sigma_1\,b\cdot\sigma_2\,|\psi> \= \- a\cdot b.
\ee

In particular, when Alice and Bob choose to measure in the same direction ($a = b$), they should observe perfect anti-correlations, 
while also observing maximal randomness (``up" and ``down" with equal probabilities). (If we had available an exact solution to \Schseqn\ for EPRB, it will
probably not contain these exact correlations; see section \ref{false_chap}.) 

Bell proved his theorem assuming the EPRB correlations would appear as QM seems to predict,
and certain assumptions about any theory of this phenomenon that satisifies a realism criterion.
For the latter he assumed a universal probability law on a ``hidden variable" $\lambda$, which he wrote as $\rho(\lambda)$, and two outcome variables with values $\pm 1$: $A(a,\lambda)$
and $B(b,\lambda)$, so that the correlations become

\be
P(a,b) \= \int\,d\lambda\,\rho(\lambda) A(a,\lambda)\,B(b,\lambda).
\ee

\ni He then demonstrated that 

\be
1 \+ P(b,c) \geq |\,P(a,b) - P(a,c)\,|,
\ee

\ni for any choices of angles $a,b,c$, a result known as ``Bell's Inequality." It is violated by the formula

\be
P(a,b) \= - a\cdot b.
\ee

However, as a statistician pointed out to this author sometime in the 1970s,  Bell's set-up is not necessary.
Statisticians construct probability models that contain ``covariates," often quantities under the control of the experimenter like $a$ and $b$ of EPRB, 
and ``outcome variables" like the two measured ``spins." But they put the
covariates in the probability law. (Physicists often do so, too, e.g., temperature in Gibb's distributions of thermodynamics does not have a ``universal law" but enters as a covariate.) 
So why hadn't Bell written $P_{a,b}(\lambda)$ and $A(\lambda)$, $B(\lambda)$?

In 1969, J.F. Clauser, R.A. Holt, A. Shimony, and M.A. Horne 
derived a more general Bell-type inequality (abbreviated CHSH), \cite{chhs}; Bell did so as well, \cite{bell2}.
In 1988, Edward Nelson published a particularly clear analysis of the assumptions necessary to derive a Bell-type inequality with the more general set-up, \cite{nelson}. 
He differentiated between two senses of ``locality" in the hidden variables: {\em active locality} and {\em passive locality}.
``Active locality" rules out action at a distance (that would imply a violation of
Einstein's relativity principle) by asserting that $P_{a,b}(A = \pm 1)$ doesn't depend on $b$ 
(Bob's choice of measurement direction does not influence the marginal probabilities for what Alice observes, and {\em vice versa}).
``Passive locality," by contrast, is a rather subtle condition reminiscent of the Markov law in the theory of stochastic processes. (The Markov law reads: 
the future and the past are independent, given the present. A common mistake with the Markov law is to postulate that the factorization holds conditional {\em on your favorite observables} in the present; 
actually, you must condition on everything random for it to apply, lest influences from the past sneak around into the future). The condition is: there exist some variables, observable or not,
in the casual past
of both labs such that, conditional on all these variables, $P_{a,b}$ factorizes as $P_a\,\times\,P_b$ (outcomes in the two labs are statistically independent, conditional on the
extra variables). For use in explicating the measurement theory presented here, I present another proof of CHSH from Nelson's axioms in the Appendix to this paper.

From the work of these physicists and mathematicians, we can derive two remarkable conclusions. First, if the QM predictions for EPRB are ever verified,
{\em local realism} is refuted. 
(For discussion of some recent experiments 
that may have demonstrated a violation of Bell's inequality in something similar to EPRB, see section \ref{false_chap}.) 
Second, any theory meant to improve on QM  
must {\em violate passive locality}.
In particular, such an alternative theory must be, in contrast to Heisenberg's or \Sch's original theories, {\em stochastic} (i.e., 
describe explicit random elements in the underlying strata of reality).

\pagebreak

\subsection{Stochastic \Sch\ theory}

The paradigm supported in these papers is \Sch's. \Schists\ do not accept the statistical interpretation of the wavefunction 
(as a catalog of correlations, or a compendium of ``our
knowledge of the particle's location and momentum"). Thus, we must seek another explanation for random outcomes in certain experiments.
From the above developments, today we know that any extension of \Sch's theory to explain measurements
will have to be {\em stochastic} and (in some sense) {\em non-local}. The only object we possess that can fulfill these requirements is the \wf\ itself. 

In the next section I will introduce a 
random component into the theory which will appear somewhat {\em ad hoc}. 
Speculation about more-detailed extensions or the origin of the randomness is relegated to later sections. 

\section{The Random Part of the Wavefunction\label{random_sect}}

I chose EPRB as the context in which to describe the stochastic part of the theory. The basic idea is that $\psi$ 
has both deterministic and random parts:

\be
\psi \= \hpsi \+ \phi,\label{psisum}
\ee

\ni where $\hpsi$ is the deterministic part and $\phi$ the random part. (As written in (\ref{psisum}), $\psi$ is not normalized to one. Multiplying by $1/\sqrt{2}$ restores normalization
in expectation over the random part but otherwise changes only the numerical interpretations of measurement thresholds, below, which is not important.) 
Both are functions of $x_1,x_2,...;t$, as usual. The deterministic part, $\hpsi$, satisfies \Schseqn\ while the random part, $\phi$,
is a centered \rv\ with mean zero to be described shortly. As I do not have an exact solution of \Schseqn\ for EPRB, I assume an idealized ``solution" of form:

\bar
\no \hpsi &\=& c_{+,+}\,\theta_{A,a,+}\,\theta_{B,b,+}\,|++> \+ c_{-,+}\,\theta_{A,a,-}\,\theta_{B,b,+}\,|-+> \+\\
 &&  c_{+,-}\,\theta_{A,a,+}\,\theta_{B,b,-}\,|+-> 
\+ c_{-,-}\,\theta_{A,a,-}\,\theta_{B,b,-}\,|-->
\ear

\ni where $\theta_{A,a,+}$ denotes a wavepacket that represents a ``particle displaced upward in direction $a$," in Alice's laboratory, and so forth. 
The coefficients $c_{..}$ are derived from solving \Schseqn\ starting with a ``two-particle, total spin zero" \wf\ and can be written: 

\be
c_{+,+} = \sqrt{p_{+,+;a,b}}\exp\left(\,i\,\gamma_{+,+;a,b}\,\right),
\ee

\ni etc., where the $p_{\cdot,\cdot}$ are the familiar probabilities:

\bar
\no p_{+,+;a,b} \= p_{-,-;a,b} &\=& (1/4)\,\left(\,1 - a\cdot b\,\right);\\
\no p_{+,-;a,b} \= p_{-,+;a,b} &\=& (1/4)\,\left(\,1 + a\cdot b\,\right).
\ear

\ni and the $\gamma_{..}$ are real phases. 

Lacking a dynamical theory of the random part, I rely simply on making it as random as possible, while assuming the same form and properties as for $\hpsi$:

\bar
\no \phi &\=& v_{+,+}\,\theta_{A,a,+}\,\theta_{B,b,+}\,|++> \+ v_{-,+}\,\theta_{A,a,-}\,\theta_{B,b,+}\,|-+> \+\\
 &&  v_{+,-}\,\theta_{A,a,+}\,\theta_{B,b,-}\,|+-> 
\+ v_{-,-}\,\theta_{A,a,-}\,\theta_{B,b,-}\,|-->
\ear

\ni where the $v_{..}$ are chosen to be ``maximally random" complex coefficients subject to two conditions:
the normalization

\be
|v_{+,+}|^2 \+ |v_{+,-}|^2 \+|v_{-,+}|^2 \+|v_{-,-}|^2 \= s^2,\label{constraint_eqn}
\ee

\ni and the symmetry condition

\def\Ps{P_{\hbox{sym.}}}
\be
\Ps\,\phi \= \phi.
\ee

\ni where $\Ps$ denotes the projection on symmetrical states (which include $\hpsi$). I will also consider an example with Gaussian $v$'s.

\section{Measurement with Random Outcomes\label{randomout_sect}}

We know from experiments (or we believe because we think QM got it right) that in EPRB there are detections 
in each laboratory which are random and correlated between laboratories.
How could this occur in NLQM with a random part of the \wf? 

I denote by $G_{A,a}$ a function of the ``particle's position" in Alice's lab which has roughly two values
$\pm1$, with $+1$ meaning ``the particle has entered the apparatus at the `up' 
side in direction $a$, and so forth. 
(If G is a smooth function it will necessarily take other values including zero. 
However we can assume the wavepackets representing ``particle went up," etc., at the moment of measurement are concentrated where G is $\pm$ 1.)
I then assume that the force on the apparatus needle is:

\be
F_A \= <\psi|\,G_{A,a}\,|\psi>
\ee

\ni and similarly for Bob's lab. 
Plugging in 
(\ref{psisum}) we can write:

\bar
\no F_A &\=& <\psi|\,G_{A,a}\,|\psi>\\
\no &\=& <\hpsi|\,G_{A,a}\,|\hpsi> \+ 2\,\Re\, <\hpsi|\,G_{A,a}\,|\phi> \+ <\phi|\,G_{A,a}\,|\phi>.\\
&& \label{forceeqn}
\ear

Now assuming that the inhomogenous magnetic fields (which are really part of the apparatus) at both labs yields cleanly separated, normalized wavepackets, 
in (\ref{forceeqn}) the first and third terms should be $\approx 0$ by symmetry and the middle term is:

\be
F_A \= 2\,\Re\,\left\{\, c_{+,+}\,v_{+,+} \+ c_{+,-}\,v_{+,-} \- c_{-,+}\,v_{-,+} \- c_{-,-}\,v_{-,-}\,\right\}.
\ee

Similarly for Bob's lab:

\be
F_B \= 2\,\Re\,\left\{\, c_{+,+}\,v_{+,+} \- c_{+,-}\,v_{+,-} \+ c_{-,+}\,v_{-,+} \- c_{-,-}\,v_{-,-}\,\right\}.
\ee

Now I assume that the apparatus are such that 

\be
\Plb \hbox{``up at A"} \Prb \= \Plb F_A > \beta \Prb.
\ee

\ni Here $\beta$ is a small positive number representing the threshold ``kick" needed to get the apparatus to register, which I discuss further below. 
Similarly,

\bar
\no \Plb \hbox{``down at A"} \Prb &\=& \Plb F_A <  - \beta \Prb;\\
\no \Plb \hbox{``up at B"} \Prb &\=& \Plb F_B > \beta \Prb;\\
\no \Plb \hbox{``down at B"} \Prb &\=& \Plb F_B < - \beta \Prb;\\
&& 
\ear

With this scheme, if $\beta > 0$ there will be some non-detection events. Non-detections and double-detections I discuss in the next section; 
so for the moment let's assume $\beta = 0$. The next issue arising: how do these signs translate into measurement outcomes? That's the job of the ``particle detector(s)." So far
in this series, as appartus I have imagined only a needle balanced at the unstable point of an external potential and which receives a ``kick" resulting in motion up or down.
(The non-linear part of the Hamiltonian is assumed to have enforced classical behavior on the apparatus, meaning that the needle cannot move in both directions simultaneously.) 
We thus obtain discrete outcome variables by
examining only signs, not magnitudes. Next question: are the probabilities for the joint outcomes implied by the above definitions, namely: 

\def\Puu{P_{\hbox{up, up}}}
\def\Pud{P_{\hbox{up, down}}}  
\def\Pdu{P_{\hbox{down, up}}}
\def\Pdd{P_{\hbox{down, down}}}

\bar
\no P_{\hbox{up, up}} &\=& \Plb F_A > 0;\,F_B > 0 \Prb;\\  
\no P_{\hbox{up, down}} &\=& \Plb F_A > 0;\,F_B < 0 \Prb;\\  
\no P_{\hbox{down, up}} &\=& \Plb F_A < 0;\,F_B > 0 \Prb;\\  
\no P_{\hbox{down, down}} &\=& \Plb F_A < 0;\,F_B < 0 \Prb;\\ 
&&
\ear

\ni the same as the QM predictions? In particular, is the correlation coefficient:

\be
C_{a,b} \= \Puu \- \Pud \- \Pdu \+ \Pdd
\ee

\ni equal to $- a.b$? Although the exact anticorrelations when $a = b$ 
are preserved, the answer is no. (See section \ref{contrast_sect}.) 
What about Bell's inequality? Although it would be possible to answer that question analytically, 
I found it easier (and probably more reliable) to
turn to computer simulations, which also make it convenient to search among pairs of angles for a maximal violation of, e.g., the CHSH inequality. Defining:

\be
S \equiv C_{a,b} \+ C_{a',b} \+ C_{a,b'} \- C_{a',b'},\label{Sdef}
\ee

\ni the boundary of local realism is $|S| \leq 2$, while the maximal violation predicted by QM is $2\sqrt{2}\approx 2.8$. 
With the ``standard angles" $a = 0;a' = 90;b = 225;b' = 315$, the model yielded $S = 2.11$; a search through 1,000 angle pairs found one with $S = 2.84$.

\section{Non- and Double-Detections\label{double_sect}}

A needle balanced at a critical point of an external potential may strike the reader as unrepresentative of real ``particle detection" apparatus. Moreover, as \Schism\ does not
entertain real ``particles" there arises the strange possibility of double detections. I discuss this issue in this section.

Let's begin by considering a single Stern-Gerlach (SG) set-up. The corresponding random \wf\ takes the form:

\bar
\no \psi &\=& \hpsi \+ \phi\\
\no &\=& \sqrt{p}\,\theta_{+}(y) \+  \sqrt{1 - p}\,e^{i\,\gamma}\,\theta_{-}(y) \+ \\
&&  v_{+}\,\theta_{+}(y) \+ v_{-}\,\theta_{-}(y).\label{sg_eqn}
\ear

\ni where $\gamma$ is a fixed real phase, the $v_{.}$ are random complex numbers, and I have written the general case for use later (SG of course has $p = 1/2$).

Next, rather than the single apparatus with the needle, let us imagine two apparatus located some distance up or down the $y$-axis, and two functions $G_1(y)$ and $G_2(y)$ which take values 0 or 1
and forces

\be
F_1 \= <\psi|G_1|\psi>; \ph\ph  F_2 \= <\psi|G_2|\psi>.
\ee

Plugging in from (\ref{sg_eqn}), these are given by:

\bar
\no F_1 &\=& p \+ 2\,\hbox{Re}\,\sqrt{p}\,v_{+} \+ |v_{+}|^2 \\
 F_2 &\=& 1 - p \+ 2\,\hbox{Re}\,\sqrt{1 -p}\,v_{-} \+ |v_{-}|^2.
\ear

\ni (Since $v_{-}$ has a random phase, the factor including $\gamma$ can be absorbed into it.)

Now assume that the apparatus has a threshold, call it $\beta > 0$, which the force must exceed before the apparatus registers. 
I abbreviate possible outcomes as: detection (one apparatus registers), D; it didn't, N;
no detections, ND; and double detection (both apparatus register), DD. Then outcome probabilities are:

\bar
\no \Plb \hbox{App. 1, D; App. 2, N} \Prb &\=& \Plb F_1 > \beta;\ph F_2 < \beta \Prb;\\
\no \Plb \hbox{App. 1, N; App. 2, D} \Prb &\=& \Plb F_1 < \beta;\ph F_2 > \beta \Prb;\\
\no \Plb \hbox{DD} \Prb &\=& \Plb F_1 > \beta;\ph F_2 > \beta \Prb;\\
\no \Plb \hbox{ND} \Prb &\=& \Plb F_1 < \beta;\ph F_2 < \beta \Prb.
\ear

In this model, a large $\beta$ will generate a lot of ND's, while a small $\beta$ will generate many DD's. So I resorted to simulations to study
these fractions as a function of two parameters: $\beta$ and $s$, the size of the random part:

\be
|v_{+}|^2 \+ |v_{-}|^2 \= s^2.
\ee

First case (here $p = 1/2$): with $\beta = s = 1$, there were 12 percent DD's and 33 percent ND's. What does this mean? ND's represent no occurrence;
normally they do not appear in the data. To discover an ND rate, one would need a third apparatus capable of detecting ``the particle is on its way!" which for photons is impossible to arrange,
and even for magnetic atoms very difficult without perturbing the ``particle's spin state," because of the backreaction phenomenon in a quantum measurement, to be discussed in section \ref{repeated_sect}.
Adding a third apparatus would require an expanded wavefunction incorporating that apparatus's variables. And there would be a threshold for the additional device; if 
this device triggered, that might increase the probability that a downstream device does, too. Finally, if the detection apparatus was actually a photographic plate
(so that there are in a sense a large number of ``detectors"), even if the added detector reported ``particle coming," it would hardly be surprising if no grain in the emulsion was hit. 
Nevertheless, this case might be ruled out by some well-designed experiment.

By contrast to ND's, DD's are surprising in any variety of \paism. How can ``the particle" go both ways? In principle, the observation of a large DD rate could refute real-particle theories
(as well as the Born-von Neumann measurement axiom).
A smaller DD rate would probably be explained away. All apparatus produce strange outcomes on occasion. Indeed, ``photon detectors" 
have a ``window-width" parameter (usually referring to a temporal window); given the beam intensity, one selects it so that ``at most one photon of the given energy" will enter the detector 
most of the time.
Set the window too narrow and you don't see anything; set it too large and you get multiple, continuous detections. 

So while a large ND rate is unlikely to be troublesome, a large DD rate in experiments might falsify this model. So I used the computer to search the region:
$0 < \beta \leq 2$, $0 < s \leq 1$ for the minimum DD rate, which occurred at $\beta = 2$ and $s = 1$ with DD rate = 0 and ND rate = 0.68. (Searching for the minimum DD + ND rate found the same 
result.)

\begin{figure}
\rotatebox{0}{\resizebox{5in}{5in}{\includegraphics{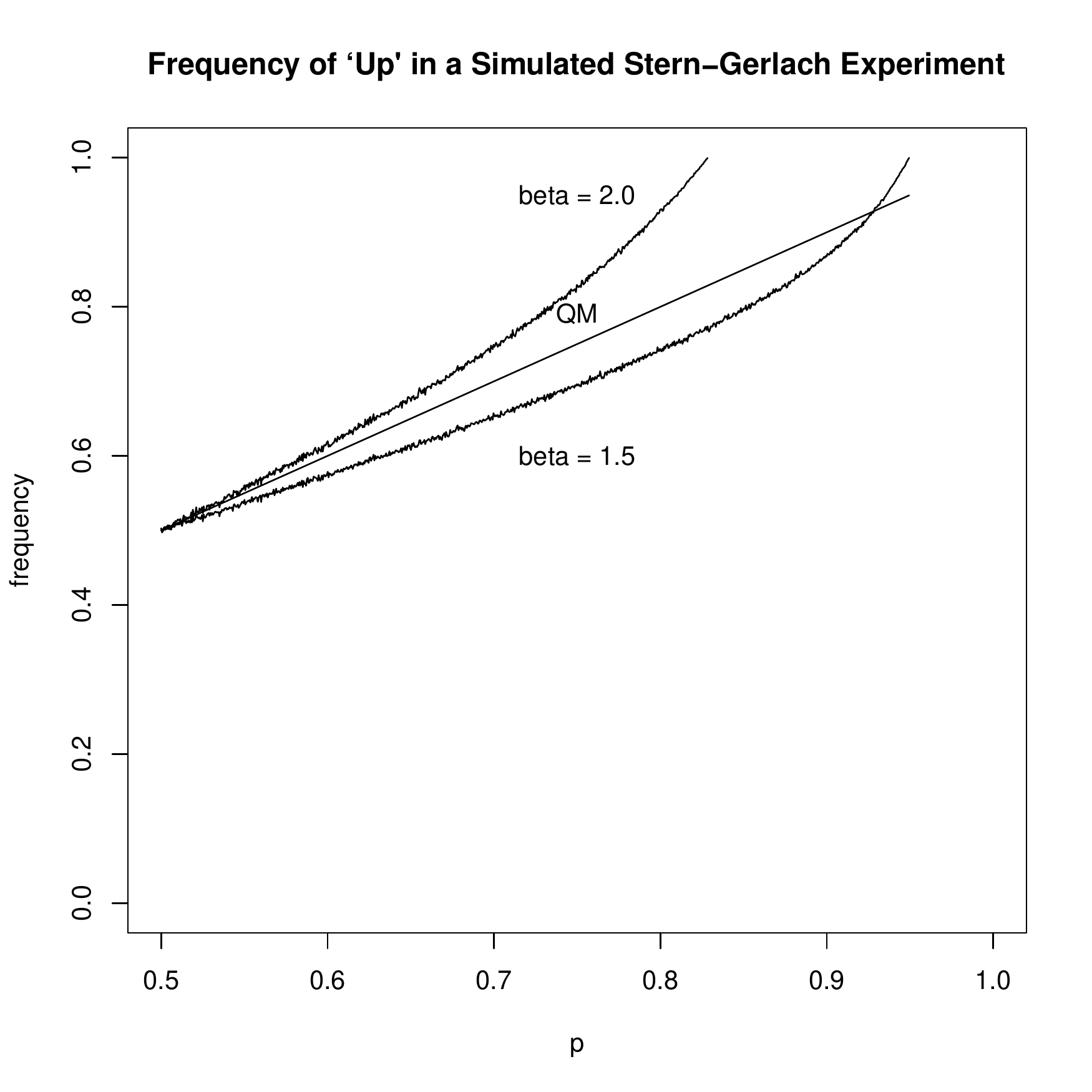}}}
\caption{Frequencies for ``up" outcomes simulated for the SG experiment. The straight line is the \vN\ assumption.\label{sgfig}}
\end{figure}

What about the cases with $p \neq 1/2$? These are interesting, assuming they can be created experimentally, because the \vN\ axiom amounts to a measurement of $p$ {\em via} frequencies. 
Figure \ref{sgfig} shows the results for two values of the threshold, $\beta$. Note that in the case $\beta = 2$, the experiment fails to resolve $p$'s for $p > 0.8$ (but has no DDs);
with $\beta = 1.5$ we get better resolution, closer to a true \vN\ measurement (but about 2.6 percent DDs appear at $p = 1/2$).

Now let us consider EPRB with the ``dual apparatus" at both ends. In addition to double- and no-detections,
we now have the additional possibility of ``single detections" (SD), meaning an event registered 
in Alice's lab but not in Bob's, or {\em vice versa}. We have four forces given by:

\bar
\no F_{A,1} &\=& 1/2 \+ 2\,\Re \, \left\{\,c_{++}\,v_{++} \+ c_{+-}\,v_{+-}\,\right\} \+ |v_{++}|^2 \+  |v_{+-}|^2; \\
\no F_{A,2} &\=& 1/2 \+ 2\,\Re \, \left\{\,c_{-+}\,v_{-+} \+ c_{--}\,v_{--}\,\right\} \+ |v_{-+}|^2 \+  |v_{--}|^2; \\
\no F_{B,1} &\=& 1/2 \+ 2\,\Re \, \left\{\,c_{++}\,v_{++} \+ c_{-+}\,v_{-+}\,\right\} \+ |v_{++}|^2 \+  |v_{-+}|^2; \\
\no F_{B,2} &\=& 1/2 \+ 2\,\Re \, \left\{\,c_{+-}\,v_{+-} \+ c_{--}\,v_{--}\,\right\} \+ |v_{+-}|^2 \+  |v_{--}|^2.\\
&\=&\label{F_eqns}
\ear

and the corresponding outcomes:

\be
\Plb \hbox{``up,up"} \Prb \= \Plb F_{A,1} > \beta; F_{A,2} < \beta; F_{B,1} > \beta; F_{B,2} < \beta \Prb,
\ee

\ni and so forth. For the case $\beta = 2.0,\ph s = 1.0$ as above I found from 100,000 simulations: for zero angle between $a$ and $b$, 
DD = 0, SD = 4.2 percent, and ND = 92 percent (with a correlation, computed using only the ``allowed" events with one detection at each lab, of exactly -1). 
With angle difference of 90 degrees, DD = 0, SD = 9.6 percent, ND = 77 percent (and correlation 0.80; this surprising
figure will be discussed in the next section). 
With ``standard angles" and the dual-apparatus, the CHSH statistic gave: $S = 3.37$.

\section{The Contrast with Conventional QM\label{contrast_sect}}

\begin{figure}
\rotatebox{0}{\resizebox{5in}{5in}{\includegraphics{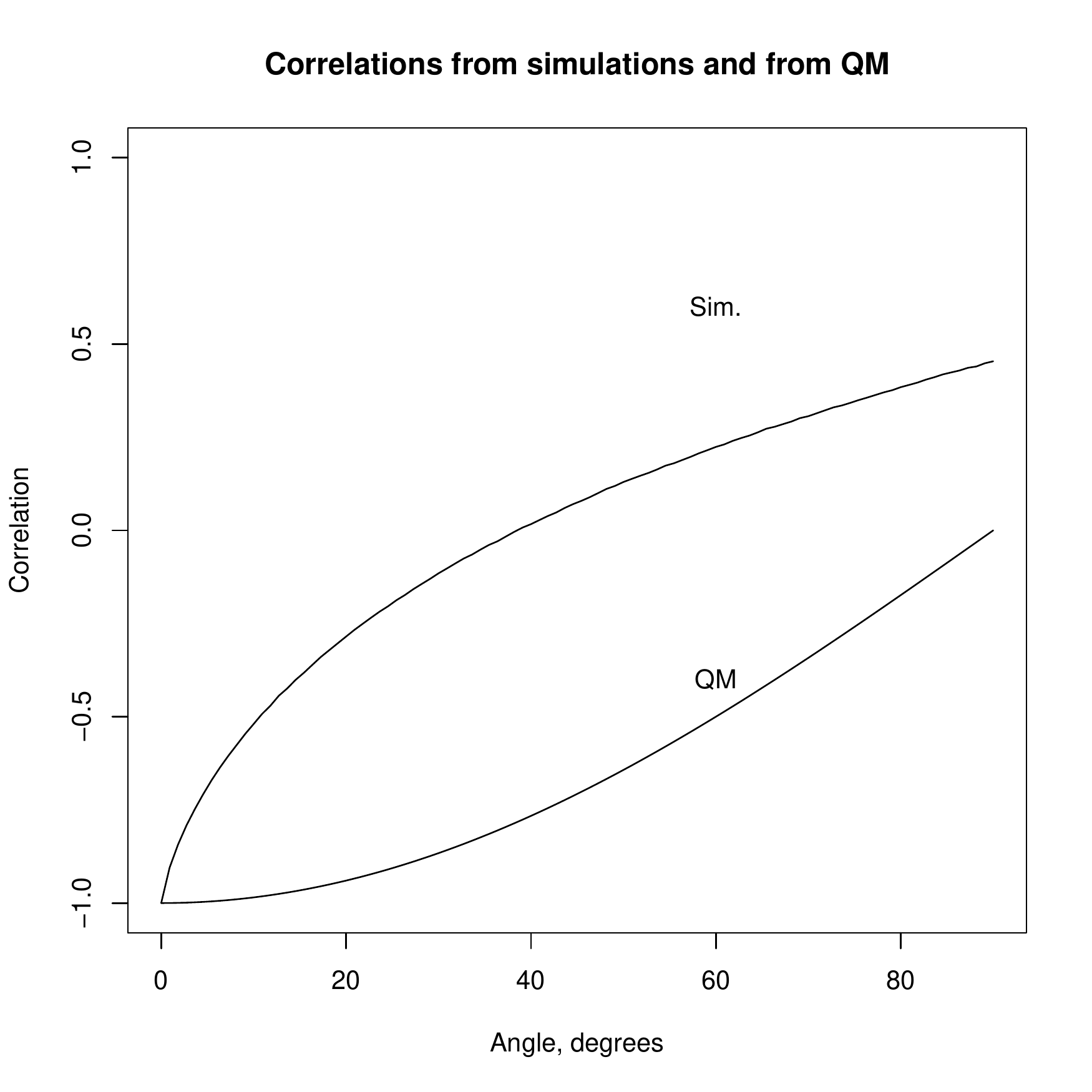}}}
\caption{Correlations from simulations of the ``single apparatus" model of EPRB, showing also the \vN\ prediction. The difference between Alice's and Bob's directions
of measurement are plotted on the x-axis.\label{corrfig1}}
\end{figure}

As I remarked in the previous section, the correlations in EPRB I found are not those of QM assuming a \vN\ measurement of the two ``spins."
Look first at Figure \ref{corrfig1}. The correlations from simulating the ``single apparatus" model of SG, used in the EPRB context, are larger than the QM prediction ($- a\cdot b$),
agreeing only at zero angular difference.

\begin{figure}
\rotatebox{0}{\resizebox{5in}{5in}{\includegraphics{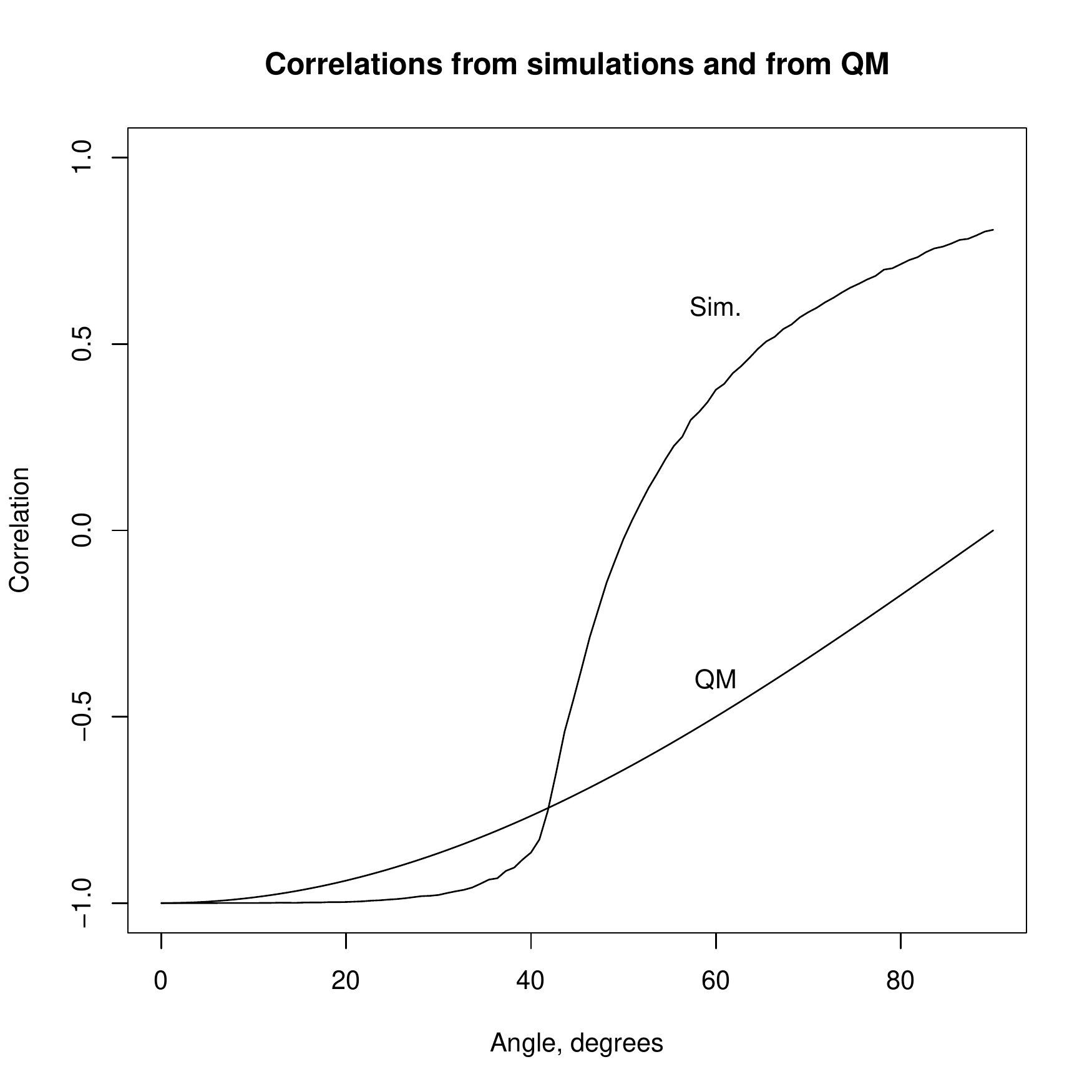}}}
\caption{Correlations from simulations of the ``dual apparatus" model of EPRB, showing also the \vN\ prediction. The difference between Alice's and Bob's directions
of measurement are plotted on the x-axis.\label{corrfig2}}
\end{figure}

\begin{figure}
\rotatebox{0}{\resizebox{5in}{5in}{\includegraphics{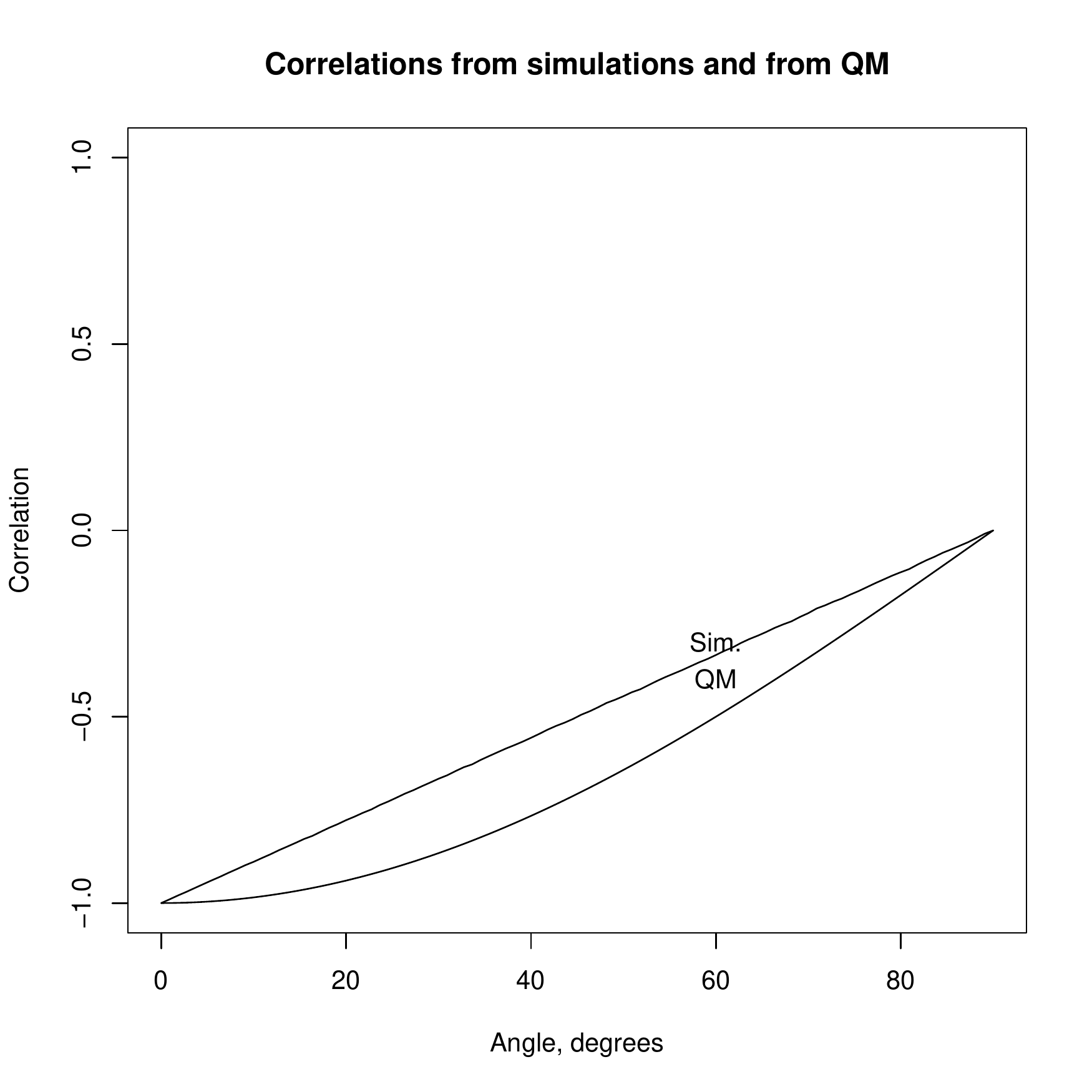}}}
\caption{Correlations from simulations of the ``single apparatus" model of EPRB with Gaussian random part, showing also the \vN\ prediction. The difference between Alice's and Bob's directions
of measurement are plotted on the x-axis.\label{corrfig3}}
\end{figure}

\begin{figure}
\rotatebox{0}{\resizebox{5in}{5in}{\includegraphics{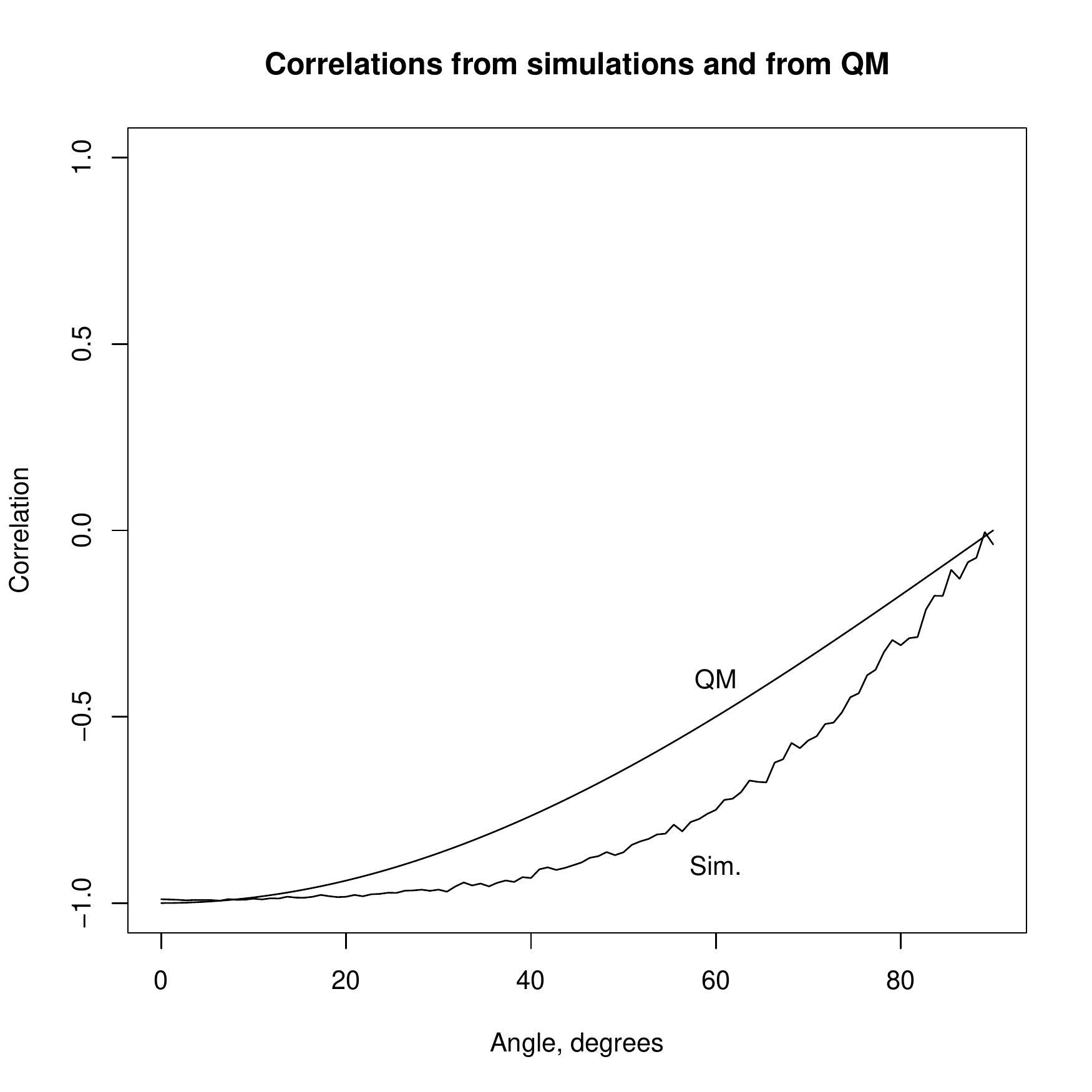}}}
\caption{Correlations from simulations of the ``dual apparatus" model of EPRB with Gaussian random part, showing also the \vN\ prediction. The difference between Alice's and Bob's directions
of measurement are plotted on the x-axis.\label{corrfig4}}
\end{figure}

Next look at Figure \ref{corrfig2}. The correlations from simulating the ``dual apparatus" model in EPRB cross over the QM prediction at about 45 degrees.

Finally, I altered the random part $\phi$ of the \wf\ by replacing bounded $v$'s by independent Gaussians of variance 0.25. See Figures \ref{corrfig3} and \ref{corrfig4}.
These curves are much closer to the predictions of standard QM. For the CHSH statistic, the single-apparatus case with Gaussian random part yielded (standard angles) $S = 1.99$ 
(a search of 1,000 choices yielded no increase),
and the dual-apparatus case, $S = 3.53$.

Clearly, these figures yield the possibility of falsifying the theory presented here. (I delay that discussion until section 7.)
\pagebreak

\section{Repeated Measurements and Back-Reaction\label{repeated_sect}}

Eventually we must address the question: what is ``the state of the measured system after the measurement?"
The salient remark is that in \Schism\ the question is meaningless, because (as \Sch\ remarked in his ``cat paper" of 1935, \cite{catpaper})
the microscopic system does not have its own wavefunction. Instead, we must add a second measuring apparatus downstream from the first and ask about the sequence of outcomes.

\begin{figure}
\rotatebox{0}{\resizebox{5in}{5in}{\includegraphics{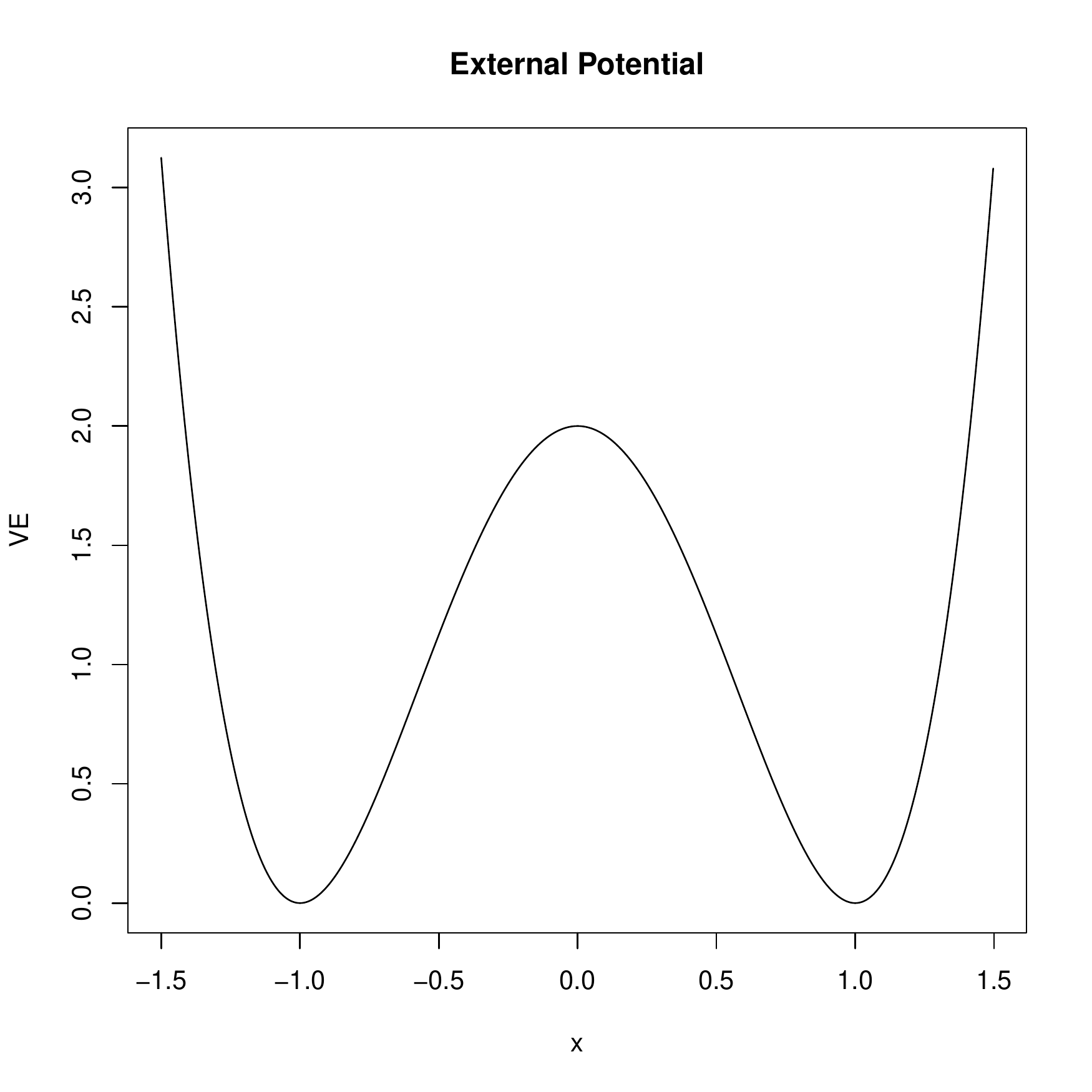}}}
\caption{Example of an unstable external potential, for use in the ``single-apparatus" SG model.\label{vefig}}
\end{figure}

Unfortunately, I cannot solve nonlinear \Sch\ equations analytically. However, by assuming that the nonlinear terms in the Hamiltonian do their job of restoring classical physics for
the apparatus pointer(s), I can say something for a highly simplified, unrealistic SG-type model of multiple ``spin" measurements. Accordingly, let $X$ as before represent the COM of the first apparatus,
which I take in the ``single-apparatus" model to be a pointer needle initially balanced at the critical point of an unstable potential written $V_E$, for example the quartic shown in Figure \ref{vefig}. 
(As usual, I assume the pointer is free to move
in a single dimension, which need not be the x-axis.) The ``particle position" as before is $y$ (think of the ``particle" moving along the x-axis, but deflected up or down the y-axis)
with $Y = <\psi|y|\psi>$. Y is not itself observerable (being microscopic), but of interest nonetheless. I take for the ``magnetic field" operator in the quantum Hamiltonian:

\be
V_M \= \- M\,y\,\sigma_y,
\ee

\ni where M stands for magnetic field
and $\sigma_y$ for the Pauli spin-operator
($\sigma_y\,|\pm> = \pm\,|\pm>$). (The minus sign before M is for convenience below.) For the interaction of macroscopic with microscopic I assume the unrealistic choice:

\be
V_I \= \alpha\,\sum_{i=1}^N\,x_i\,y.
\ee

\ni where $\alpha$ is a positive constant. 

Taking two time-derivatives then yields (calculation left to the reader; assume that the interchange of sums in the $V_E$-term is permissible because of the NL terms):

\bar
\no \ddot{X} &\=& - V_E'(X) \- \alpha\,Y;\\
\no \ddot{Y} &\=& - \alpha\,N\,X \+ \\
\no &&  M\,\left\{\,(2\,p - 1) + 2\,\Re\,\left(\, \sqrt{p}\,v_{+} - \sqrt{1-p}\,v_{-}\,\right) + |v_{+}|^2 - |v_{-}|^2\,\right\}.\\
&&\label{macroeqns}
\ear

For initial conditions take: $X=0$; $\dot{X}=0$; $Y=0$; $\dot{Y}=0$. (I have omitted some factors of $N$ and $m$ 
because I am going to graph in arbitrary units.) Now suppose the random force (containing the $v$'s) happens to be positive. Then there will be a positive force on $Y$,
which will begin to move in the positive direction, which will produce a negative force on X (increased, with same sign, by the external potential), 
which will begin to move in a negative direction, which in turn puts more positive force on Y, and so forth.
The needle will move down. We should nevertheless label that outcome an ``up" detection. And conversely if the random force is negative (the pointer goes up, which we label a ``down"). 

Note the factor of $N$ in the second equation, first term on the right, of (\ref{macroeqns}). Even if $\alpha$ is very small, there will be a substantial ``kick" on the microsystem, in the opposite
direction from which the pointer moves. This is the well-known ``back reaction" characteristic of quantum measurements, which distinguishes them from classical measurements.

\begin{figure}
\rotatebox{0}{\resizebox{5in}{5in}{\includegraphics{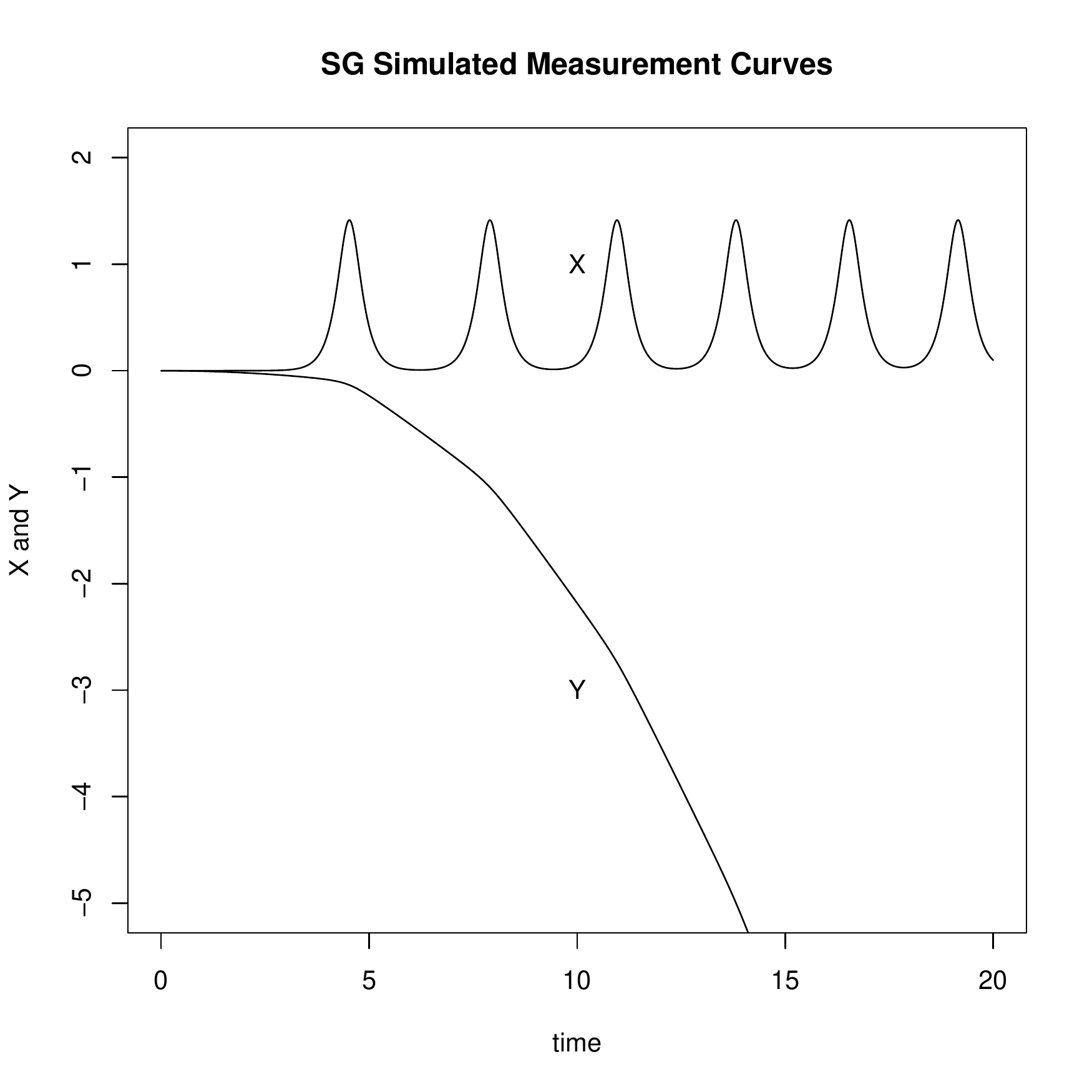}}}
\caption{Simulated curves of the SG measurement model, with a small negative force on Y.(Arbitrary units on the y-axis.)\label{sympfig}}
\end{figure}

\begin{figure}
\rotatebox{0}{\resizebox{5in}{5in}{\includegraphics{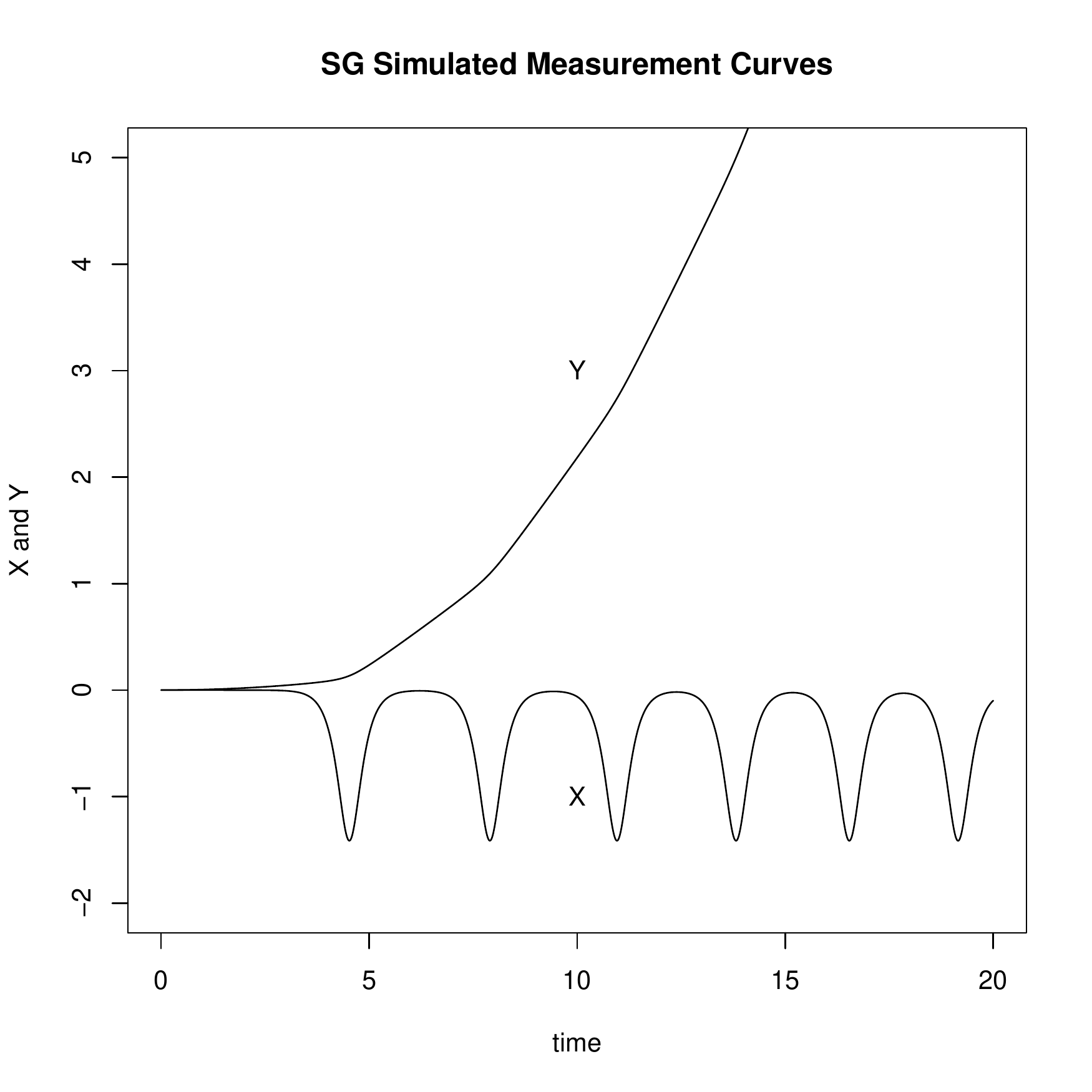}}}
\caption{Simulated curves of the SG measurement model, with a small positive force on Y. (Arbitrary units on the y-axis.)\label{sympfig1}}
\end{figure}

It is not difficult to simulate the solutions of (\ref{macroeqns}) on the computer. (I used a symplectic solver; see Computational Appendix.) Figures \ref{sympfig} and \ref{sympfig1}
show two representative curves. The apparatus COM in each settles into a stable oscillation around one of the stable points of $V_E$; Y moves in the opposite direction.

\def\secX{X_{\hbox{sec.}}}

Now introduce a second apparatus with COM $\secX$. Assuming the ``particle" has to travel some distance along the x-axis before encountering the second apparatus, we need only write
another equation for it:

\be
 \ddot{X}_{\hbox{sec.}} \= \- V_E'(\secX) \- \alpha\,Y,
\ee

\ni and add a term: $- \alpha\,N\,\secX$ to the right side of the equation for $\ddot{Y}$. 
But now we take initial conditions:$X_{\hbox{sec.}}=0$; $\dot{X}_{\hbox{sec.}}=0$; $Y=Y(T)$; $\dot{Y}= \dot{Y}(T)$, where $T$ is some later time when the ``particle reached the second apparatus." 
Now we see that, if $Y(T)$ is positive, there will be a negative
force on $\secX$. Thus we are going to get a second pointer moving down and the overall result is ``up, up." The only other possibility is ``down, down." 
Hence the illusion that ``a particle (in a definite spin state)
travelled from the source through the two apparatus in succession." (In 1929, Nevill Mott made a similar argument to explain those ``particle tracks" seen in cloud chambers 
on the basis of wave mechanics, \cite{mott}.)

This scenario brings up an issue about repetition of the random force. I assumed it did not change between the measurements. 
The other candidate hypothesis is that, during the time between measurements, there occurred what probabilists call ``innovations:" independent random effects on the ``particle's spin." 
Whether we would then get, in the double-y-measurements,
``up, down" or ``down, up" would depend on the relative magnitudes of forces (i.e., would involve the assumed values for $\alpha$, $N$, and $M$). Given the large
``back reaction" effect from the first measurement, many choices of these parameters would suppress a scenario in which an opposite random force succeeded in reversing
the outcome. 
\dotskip

What about sequential measurements of ``non-commuting observables?" Von Neumann's measurement axioms can be amplified a bit to include the statements: (a)
simultaneous measurement of two non-commuting observables (Hermitian operators)
is impossible;
and (b), sequential is possible 
and the probabilities for the second measurement are derived from the ``collapsed" state vector given the outcome of the first. 
The set-up of this section agrees with the first statement in that the appparatus needle was assumed to move only in the y-direction, making simultaneous observation of the ``z-component of spin" impossible.
So let us ask what happens if the second measurement occurs at some time $T$ after the first (assumed non-destructive in the sense that ``the particle flies on"), with a second apparatus
needle that moves in the z-direction. Let's assume that the motion of ``the particle" in the y-direction is irrelevent for the second measurement in the z-direction (imagine two z-measuring apparatus
placed some distance up and down the y-axis, where two ``beams" are directed by the first apparatus magnetic field). 
The set-up for $X_{\hbox{sec}}$ and $Z$ starting at time $T$ is then essentially the same as above for $X$ and $Y$ starting at time zero, 
except that the random part of the force on $Z$ is replaced by $M$ times:

\be
<\psi|\sigma_z|\psi> \= 1 \+ \sqrt{2}\,\Re\,\left\{\,v_{+}^* + v_{-}\,\right\} \+ 2\,\Re\,\left\{\,v_{+}^*\,v_{-}\,\right\}.\label{randzeqn}
\ee

Concerning the repetition-of-randomness issue, assuming independence (i.e., replacing $v$'s by $v'$s in (\ref{randzeqn}) with the two sets having independent distributions) would imply that the z-outcomes 
``up" or ``down" occur with equal probabilities, independent of the outcome of the y-measurement. That agrees with conventional QM; recalling that $|\pm>$ 
stood for eigenvectors of $\sigma_y$, equal probabilities follow from the calculation: $<+|\sigma_z|+> = <-|\sigma_z|-> = 0$. By comparison, 
the expression in(\ref{macroeqns}) 
second equation, term containing $M$, derived from (set $p=1/2$):

\be
<\psi|\sigma_y|\psi> \= \sqrt{2}\,\Re\,\left\{\,v_{+} - v_{-}\,\right\} \+ |v_{+}|^2 - |v_{-}|^2.\label{randzeqn2}
\ee

\ni With repeated ``maximally random" $v$'s, 
(\ref{randzeqn}) and (\ref{randzeqn2}) are orthogonal. If the $v$'s were Gaussian that would yield independence, but not in the bounded case, as we saw before;
therefore the joint probabilities of the two measurements will differ somewhat from the \vN\ predictions.

One could ask about a second measurement at an intermediate angle in the y-z plane. However, the ``back reaction" part reappears: the 
initial conditions at time $T$ would involve the motion before that time projected onto the new axis. I will forgo this scenario which would require about a doubling of the programming efforts here;
besides, this section was not intended as serious modeling of real quantum experiments (which certainly would not involve needles moving in various directions).

\section{Falsifications\label{false_chap}}

In the 1930s, philosopher of science Karl Popper pointed out that good theories at least admit the possibility of experimental falsification.
So let us ask whether the theory presented here can be, or already is, falsified.
 
The Figures showing correlations in EPRB in section \ref{contrast_sect} present one possibility. 
But that they make a {\em prima facie} case against the present theory is not immediately obvious. Recall that the real Stern-Gerlach experiment did not produce two spots,
but rather a lip-print, \cite{sg}. 
That is, there was considerable angular spread in where the ``particles impacted the screen." I have searched for authors who published an exact solution
of \Schseqn\ for SG and found two groups, but neither claimed to reproduce the lip-print with its angular dependence. (Bohm in his 1951 book did of course 
treat what came to be called the EPRB experiment, but he omitted the kinetic energy term and his ``magnetic" field did not have divergence zero, 
which it must to satisfy Maxwell's equations. 
M.O. Scully, W.E. Lamb and A. Barut, \cite{slb} 
published an expression for the SG \wf, 
but then dropped the interpretation, writing that they would return to the implications in a later publication which I did not find. Perhaps the lip-print can be derived from their formula.
M. Gondran and A. Gondran uploaded a paper to arxiv in 2005, \cite{GandG}, which displayed in one figure a bimodal density distribution, but only along the axis of measurement.) 
Clearly, one would have to make some arbitrary assumptions in order to regain the $\pm 1$-valued variables assumed in most Bell-type analyses. 
Should we put a $+1$ if the ``particle is detected above the y-axis," or only if it enters into some wedge in the upper half plane, with a specified angular width? 
(In the former case, you could have made the same declaration if an atom actually were a classical magnet with a random orientation---but nobody would agree that you had thereby
discovered the ``discreteness of the quantum world." In the latter case, if the width was narrow there would be many non-detections.) 
If someone solved \Schseqn\ exactly for EPRB and used the formulas to model a real experiment,
I doubt it would predict the exact \vN\ correlations; presumably, something like what we saw in section \ref{contrast_sect} would appear.

On the other hand, many experiments with ``photons" (with polarization replacing ``spin") from the 1970s and '80s 
yielded data apparently in perfect agreement with the QM curve of correlations. However, Emilio Santos and others pointed out that
the \nai\ QM treatment of photons from atomic cascades detected with ``photon counters" (as though another version of EPRB) is not correct, 
and Santos actually constructed a local hidden-variable model for these experiments, \cite{santos}.
Because of low counter efficiencies, experiments with photons were also vulnerable to an ensemble-selection loophole. (However, 
as I write two groups have claimed selection-loophole-free tests of Bell's inequality with photons 
and very small p-values, \cite{shalm}, \cite{guistina}.) 

Another interesting instance of the failure of QM assuming \vN\ appeared in 2015, with the publication in Nature of a reputed ``Experimental 
loophole-free violation of a Bell inequality", \cite{natpub}
The experimenters used a set-up different from EPRB that had been proposed in 1993 and shown capable of violating a Bell-type inequality. The authors introduced the subject with the usual
\vN-axiomatic analysis, leading to the conclusion that the CHSH quantity ``S" can
attain the value $2\,\sqrt{2} \approx 2.828$, where $S\leq 2$ is the boundary of local realism. Their best observed result was $S = 2.42 \pm 0.2$, from which they concluded that (finally!) 
local realism had been refuted---but, by the same criterion, it is also inconsistent with at least \vN's simple version of measurements in
quantum mechanics. (I note that an outcome obtained here in the single-detector model with ``standard angles", $S = 2.11$, lies less than two standard deviations from their result; see second line 
after eq. \ref{Sdef}.)

As I write, this experiment has been subjected to serious criticisms. 
Because the ``event-ready" detector only triggered in one ``run" in a billion,
the experimenters had only about 250 events to analyze, and had to justify publication on the basis of a p-value of .04 for violating local causality. 
But ``significance testing" in the other sciences---biology, medicine, and the social sciences---his been attacked for permitting 
too many false correlations to be reported in the literature, resulting in a ``replicability crisis" in some fields. 
By contrast, the high-energy physics community has adopted the ``5-$\sigma$" criterion,
which is much more restrictive (a p-value of .0000001).
Moreover, the loophole-free part has been questioned: A. Bednorz discovered evidence for faster-than-light signalling in the raw data, \cite{bednorz}.
Before agreeing that local realism has been experimentally refuted, we must wait for the authors to report more data. 

These examples suggest that the Born-\vN\ axioms do not have to be taken as scripture. 
We are only obliged to test predictions against data from real experiments, assuming a model for the 
apparatus employed. 
Taking into account that last sentence, double-detection (DD, e.g., in SG) or single-detection (SD, in EPRB) rates different than predicted in the model could falsify it. 
(However, as pointed out above, too many DDs would also falsify \partist\ theories.)

What does the addition of randomness to the \wf\ do to observation of that quintessential ``quantum" phenomenon, interference? To observe interference, say 
dark bands on a screen in a two-slit type experiment, requires the accumulation of many events. Nodes are present in the deterministic part of $\psi$;  
with a large average, mean-zero terms cancel out. However, with the random set-up there is an additional term not of mean zero.
Suppose that the solution of \Schseqn\ can be written: 

\be
\hpsi = c_1\,\hpsi_1 \+ c_2\,\hpsi_2,
\ee

\ni (think of $\hpsi_1$ as representing that part of the \wf\ that passed through the left slit, and $\hpsi_2$ through the right); the interference term is then:

\be
I \= 2\,\Re\,\left\{c_1^*\,c_2\,\hpsi_1^*\,\hpsi_2\,\right\}.
\ee

\def\xnode{x_{\hbox{node}}}
Suppose that, at some point $\xnode$, $I$ is negative and exactly cancels against the terms 

\be
|c_1\,\hpsi_1|^2(\xnode) + |c_2\,\hpsi_2|^2(\xnode). 
\ee

\ni If we write for two-slit, in similar fashion as I did for EPRB,

\be
\psi = \hpsi + \phi \ph\ph\hbox{and}\ph\ph \phi \= v_1\,\hpsi_1 \+ v_2\,\hpsi_2,
\ee

\ni with ``maximally random" $v$'s, and expand $\psi^*\,\psi(\xnode)$, all terms are zero or have zero mean except for:

\be
|v_1\,\hpsi_1|^2(\xnode) + |v_2\,\hpsi_2|^2(\xnode).
\ee

Whether this term can eliminate the node depends on the parameter I wrote as `$s$': the magnitude of the random part of the \wf. 
If $s$ can be taken small,
dark bands at least would not go bright. 
But there is clearly the possibility that a random part of $\psi$ could be detected by
an observed cancelling or meandering about of nodes in repeated interference experiments, or the absence of such could falsify the proposed theory. It is tempting 
to finesse the node problem by writing $\phi(x) = \hpsi(x)\,\exp[i\,\gamma(x)]$, where $\gamma(x)$ is a random phase. 
But this choice is not equivalent to the ``maximally random" case adopted above, and doesn't reproduce
the results there.

Can $s$ be small? For the single-apparatus model of EPRB,
where the force has expectation zero, the important quantity is actually $\beta/s$, meaning that, if $\beta$ is small, $s$ can be as well. 
(I chose $\beta = 0$ in section \ref{randomout_sect}, so $s$ was irrelevant and could be infinitesimal, but it's not realistic. A small $\beta$ would reflect a large amplification factor;
in the pointer-in-an-unstable-potential model, a steep and narrow potential near its local maximum would suffice.) In the dual-apparatus case, the issue is more problematic. 
With the set-up in section \ref{double_sect}, $s$ had to be substantial. (I often used $s = 1$ for the diagrams). 
However, at least in the case  of SG ($p = 1/2$), $|\beta - 1/2|/s$ is the relevant quantity, so choosing $\beta \approx 1/2$,
one can take $s$ small as well. However, such a maneuver is suspect on methodological grounds; in all forms of modeling, 
we are cautioned against creating ``knife-edge" theories---that is, theories which work only if parameters are chosen with certain critical values. 
Moreover, examining the double-detection (DD) case with $\beta = 1/2$ and $s$ very small, the condition becomes approximately: $\Re\,v_{+} > 0,\Re\,v_{-} > 0$, 
which, with random independent phases, has probability 1/4.
Perhaps experiments would then falsify the model (or all those DD's would falsify \partism).
Finally, as was mentioned in paper I of this series, it is possible that the non-linear wavefunction theory is ``chaotic", in which random outcomes might appear with exceedingly small uncertainty in
initial conditions or parameters of measurement.

\section{Discussion\label{stospec_sect}}

Why do the curves from the stochastic theory in the first form (with bounded $v$'s) in the last section show such large positive correlations near 90 degrees? That is due to the
normalization condition in (\ref{constraint_eqn}). With Gaussians the correlations are much smaller and closer to the familiar curve from QM. (The reason the correlations reach zero
at 90 degrees is that the forces, which are \rvs, are orthogonal, and for Gaussians orthogonal means independent.) 

How does this theory manage to violate a Bell's inequality? In fact, this theory is {\em actively local} but not {\em passively local}.
For active locality, recall equation (\ref{forceeqn}) or
the first two of equations (\ref{F_eqns}).
The first term in these equations (containing only $\hpsi$, the solution of \Schseqn) does not depend on Bob's choice of $b$. (Recall that  
for conventional quantum mechanics, $H_{\hbox{QM}} = H_{\hbox{A}} + H_{\hbox{B}}$ and that $H_{\hbox{B}}$ commutes 
with any operator constructed out of variables referring only to ``particle 1" and
laboratory A. This is why QM, regarded as a probabilistic theory, is itself ``actively local.") 
In the case of bounded ``maximally random" $v$'s, the second term has a universal uniform distribution. (The quantity $<\xi|\phi>$ has the same distribution 
independent of $\xi$. The symmetrization condition on $\phi$ can be dropped because $\hpsi$ satisfies it.) The third term also has a uniform distribution. 
In the Gaussian case, the same is true, because the middle terms are mean-zero Gaussians of variance one and the third term is $\chi$-squared. 

But there is a sense (somewhat subtle) in which the theory violates passive locality. (For details, see Appendix rederiving the CHSH inequality.) 
The random part has the familiar form of a \wf\ and so cannot be said to be localized to either laboratory or to a ``particle emitter"
in the past of those laboratories; nor can it be directly observed at either place. If the reader objects to postulating an entity like $\phi$ on the grounds of its nonlocal and unobservable character,
that reader should have rejected \Schism, as I defined it, at the outset. For \Schs\ \wf\ is equally nonlocal, 
and unobservable in detail.

Why did the single-apparatus, Gaussian case fail to violate Bell's inequality? This was expected, because in this case the forces are Gaussian. One way to understand
Bell's 1964 theorem is that he provided necessary and sufficient conditions for the QM two-event correlations, $- a\cdot b$, to extend to a probability distribution for
all joint outcomes of, say, observations at three angles at each lab. But for Gaussians there are no extra conditions (once the covariance of pairs is given, a distribution is defined).
Therefore, the ``discreteness" (but boundedness suffices) of quantum outcomes played a crucial role in Bell's 1964.
If you are trying to invent a realist alternative to QM, starting with Gaussians and hoping to obtain discrete outcomes by, say, 
looking only at signs, it shouldn't work. So I was somewhat surprised that the dual-apparatus, Gaussian case
violated Bell's. Perhaps that the forces had $\chi$-squared distributions rather than Gaussians played a role here.

At first sight, the random forces that act on the apparatus and determine the outcomes
appear to contain some form of action at a distance and so seem incompatible with relativity. However, reflect on the situation in conventional QM. 
If one possessed an exact solution of \Schseqn\ for EPRB,
one would note a dependence of the coefficients on $a$ and $b$. 
QM deals with this problem by asserting that the \wf\ is not itself observable; only certain quadratic functions of $\psi$,
pertaining to their respective laboratories, are observables. 
The theory presented here is similar in structure; I have not assumed that $\psi$, or $\phi$, (the random part of $\psi$), are observable,
but only the motion of, e.g., an apparatus pointer, which displayed a similar ``active locality." 

The case of EPRB with perfect alignment of the two apparatus and needles moving in opposite directions 
seemingly represents an obstruction to a relativistic theory of the type presented in these papers.
As pointed out in paper I, the nonlinear energy could vanish in this case, but for macroscopic apparatus this is a ``knife-edge" scenario 
which could not be arranged (as any discrepency from identical apparatus initial conditions or device alignments would yield $N^2$ energy scaling as before). 
However, if we imagine shrinking the apparatus in search of the classical/quantum boundary, just on the quantum side cats may form if the external potential 
can supply the requisite energy. Conceivably, in an EPRB-like situation, if Bob knows Alice's setting, he could manipulate the presence or absence of the cat state in her laboratory, 
violating active locality (and, hence, relativity).
Although this objection may be relevant to the deterministic theory, it may not be to the stochastic theory. 
The random part of the wavefunction will likely render highly improbable the exact cancellation of two large energies resulting from MD of each needle separately.
(The measurement scenarios yielding perfect anti-correlations at $a = b$ shown in the Figures of section \ref{contrast_sect} were of course highly idealized.)
Alternatively, the putative relativistic extension may contain some more localized form of the nonlinear energy, as its primary function is to
prevent macroscopic dispersion of either device (with the microsystem providing the correlations). 
At least one more parameter would be then be required.

What might we imagine for the origin of the random part of $\psi$? Here the bounded case is more amenable to speculation. 
For, e.g., EPRB, one can imagine that initially

\be
\phi \= U\,\xi,
\ee

\ni where $U$ is a random unitary operator acting on the ``spin variables" and $\xi$ is the initial ``zero total spin" state ($\xi = 1/\sqrt{2}\,|+-> \- 1/\sqrt{2}\,|-+>$). 
Some, but not all, unitary operators can be written

\be
U \= \exp\left(\,i\,J\,\delta t/\hbar\,\right),
\ee

\ni where $J$ denotes a random Hamiltonian and $\delta t$ is perhaps some time interval (from ``particle emission to detection") during which the usual quantum evolution occurs as well. 
Note that, for EPRB, 
we cannot expect factoring of form: $U = U_A\otimes U_B$, equivalently,
$J = J_A + J_B$, since to generate the maximally-random $v's$, $U$ must be chosen from a maximally-random distribution of 4x4 unitaries. 
Also, this form would not lead to a violation of Bell's inequality. 

I note that a random, symmetric Hamiltonian has been invoked in another context: justifying statistical mechanics, and thermodynamic behavior, in quantum systems, \cite{Deutsch}.

Could such a random field be consistent with the Relativity Principle? 
Consider potentials like the $\Phi_{\mu}$ of Maxwell's theory. For the expected values we can postulate something like his equations; but in a stochastic
theory one must also specify the covariances: 

\be
{\cal E}\,\Phi_{\mu}\,\Phi_{\nu} \- {\cal E}\,\Phi_{\mu}\,{\cal E}\,\Phi_{\nu} = C_{\mu,\nu},
\ee

\ni (${\cal E}$ stands for mathematical expectation) which would suffice if the potentials were Gaussian random variables. 
The matrix $C_{\mu,\nu}$ would have to be a tensor. This does not appear impossible, although it would require
inserting new constants into the theory.  

The random field is a ``contextual" introduction of randomness. A ``non-contextual" possibility would be to introduce a stochastic differential equation, say of Stratonovich form 
(in order to preserve energy on average), for the \wf. 
Again, such a theory would require more free parameters, as it would be necessary to specify
the covariances of the ``noise" increments (including decay of correlations in space and time). On the other hand, such a theory would be more definite than {\em ad hoc} or contextual randomness,
and hence more falsifiable.

\section{Appendix: A Bell-type Inequality from Nelson's Axioms \label{inequality_chap}}

\def\Pab{P_{a,b}}
\def\Paa{P_{a,a}}
\def\Eab{E_{a,b}}
\def\Eaa{E_{a,a}}
\def\Ebb{E_{b,b}}
\def\Cab{C_{a,b}}
\def\Capb{C_{a',b}}
\def\Cabp{C_{a,b'}}
\def\Capbp{C_{a',b'}}
\def\tP{\tilde{P}}
\def\Om{\Omega}

In order to explicate the exact reason why a Bell-type inequality is violated by the theory presented here, I rederive the CHSH inequality \cite{chhs}
from Edward Nelson's set-up, \cite{nelson}.
Accordingly consider a probabilistic theory with probability function $\Pab$ and
corresponding expectation $\Eab$, where $a$ and $b$ respresent controllable parameters of apparatus in Alice's and Bob's lab, respectively, 
and two random variables $\xi$ and $\eta$ respresenting outcomes at Alice's and Bob's lab, both bounded by one: $|\xi| \leq 1$ and $|\eta| \leq 1$.

Let $\Om$ stand for some existing random events in the joint casual past of Alice's and Bob's laboratories, whether observed or not. Nelson defined the theory to be {\em passively local} if, conditional
on $\Om$, random events in the two labs are statistically independent. He defined the theory to be {\em actively local} if the marginal distribution of $\Pab$ restricted to observables
in Alice's lab does not depend on anything controllable by Bob, and {\em vice versa}; moreover, the probability distribution of $\Om$ doesn't depend on the controllable quantities. (Because
$\Om$ represents something in the past of both labs. If in some sense $a$ and $b$ are random, this might fail, because dependence of either on $\Om$ might mean dependence of $\Om$ on them.
This is ruled out by the existence of free will for Alice and Bob, something we are loath to doubt.)  
Assuming $\xi$ and $\eta$ are centered r.v.'s, i.e., have mean zero:

\be
\Eab\left(\,\xi\,\right) \= \Eab\left(\,\eta\,\right)\= 0,
\ee

\ni define the correlations between measurements in the two labs by:

\be
\Cab \= \Eab\left(\,\xi\,\eta\,\right).
\ee

  With this set up we have the following:

\begin{quote}
{\bf Theorem}. Assuming active and passive locality and the other assumptions, for any four angles $a,b,a',b'$:

\be
\Cab \+ \Cabp \+ \Capb \- \Capbp \leq 2.
\ee
\end{quote}

{\bf Proof}. We have:

\bar
\no \Cab &\=& \int\,\Pab\left(\,d\Om\,\right)\, \Eab\left(\,\xi\,\eta\,|\,\Om\,\right) \\
\no  &\=& \int\,\Pab\left(\,d\Om\,\right)\,\Eab\left(\,\xi\,|\,\Om\,\right)\,\Eab\left(\,\eta\,|\,\Om\,\right)\\
 \label{dubious_eqn} &\=& \int\,\Pab\left(\,d\Om\,\right)\,\Eaa\left(\,\xi\,|\,\Om\,\right)\,\Ebb\left(\,\eta\,|\,\Om\,\right) 
\ear

\ni where the first equality follows from passive locality and the second from active locality (but see comment below). Again by active locality,

\be
\Pab\left(\,d\Om\,\right) \= \tP\left(\,d\Om\,\right),
\ee

\ni where $\tP$ is a probability law independent of $a$ and $b$. Hence, writing

\bar
\no \Lambda_a\left(\,\Om\,\right) &\=& \Eaa\left(\,\xi\,|\,\Om\,\right);\\
\no \Lambda_b\left(\,\Om\,\right) &\=& \Ebb\left(\,\eta\,|\,\Om\,\right);
\ear

\ni we have $|\Lambda_a| \leq 1$, $|\Lambda_b| \leq 1$, and

\be
\Cab \= \int\, \tP\left(\,d\Om\,\right) \,\Lambda_a\left(\,\Om\,\right)\,\Lambda_b\left(\,\Om\,\right).
\ee

Clearly,

\bar
\no |\,\Cab \pm \Cabp\,| &\=& |\,\int\, \tP\left(\,d\Om\,\right)\,\Lambda_a\,\left[\,\Lambda_b \pm \Lambda_{b'}\,\right]\,| \\
\no &\leq& \int\, \tP\left(\,d\Om\,\right)\,|\,\Lambda_b \pm \Lambda_{b'}\,| 
\ear

\ni and similarily

\be
|\,\Capb \mp \Capbp\,| \leq \int\, \tP\left(\,d\Om\,\right)\,|\,\Lambda_b \mp \Lambda_{b'}\,|.
\ee

Therefore, since for any two numbers $c$ and $c'$ with $-1 \leq c,c' \leq 1$,

\be
|c \pm c'| + |c \mp c'| \leq 2,
\ee

we obtain:

\be
|\Cab \pm \Cabp | + |\Capb \mp \Capbp | \leq 2,
\ee

\ni which contains the result. QED.
\vskip0.2in

How did the stochastic theory presented in this article---except for the Gaussian case---manage to evade this theorem? The reason is a bit subtle.
In fact, one line of the theorem may not be a consequence of the assumptions. I refer to equation (\ref{dubious_eqn}), which was said to follow from active locality.
Strictly speaking, active locality is a statement
about the unconditional distribution of observables. Thus active locality implies that
\def\Xlb{\left[\,}

\be
\Pab\Xlb \xi = \kappa \Prb = \Paa \Xlb \xi = \kappa \Prb,
\ee

\ni but it is possible that, for some values of the ``hidden variable" $\Om$

\be 
\Pab\Xlb \xi = \kappa\,|\,\Om \Prb \neq \Paa \Xlb \xi = \kappa \,|\,\Om\Prb.\label{xx_eqn}
\ee

Usually one might deny (\ref{xx_eqn}), reasoning that

\be
 \Pab\Xlb \xi = \kappa\,|\,\Om = \omega \Prb \=
\Pab\Xlb \xi = \kappa;\,\Om = \omega \Prb/\tP\Xlb \Om = \omega \Prb,
\ee

\ni and that both $\xi$ and $\Om$ are variables available, in some sense, at Alice's laboratory, so by active locality, the above conditional probability should be independent of $b$. 
But in the case considered here, this doesn't hold. Recall that Bell wrote $A(a,\lambda)$ and $B(b,\lambda)$ for his observables, with $\lambda$ standing for the ``hidden variables(s)." 
But we have instead (for the single-apparatus model), if we regard the $v_{..}$ as constituting what is denoted above by $\Om$:

\bar
\no A(a,b;\Om) &\=& 1\left[\, F_A(a,b;v_{..}) > 0\,\right] \- 1\left[\, F_A(a,b;v_{..}) < 0\,\right];\\
\no B(a,b;\Om) &\=& 1\left[\, F_B(a,b;v_{..}) > 0\,\right]\- 1\left[\, F_B(a,b;v_{..}) < 0\,\right],
\ear

\ni ($1[\cdot]$ stands for indicator function.)
It is not true that the forces are functions of only one controllable parameter. How then is active locality holding? Because 
unconditional probabilities are averages over the conditional probabilities:

\be
\Pab\Xlb \xi = \kappa \Prb = \int \tP(d\Om)\,\Pab \Xlb \xi = \kappa\,|\,\Om \Prb,
\ee

\ni and the dependence on $b$ on the left side may disappear. In fact, the {\em probability distribution} of the force $F_A(a,b;v_{..})$, regarded as a  \rv, is independent of $b$, although not functionally.

The issue, then, is whether the $v_{..}$ are observable at either lab, but we have not assumed so; they are literally ``hidden variables." 
(The only observables for the present theory
are the positions of macroscopic pointers.)
Since the outcomes are completely fixed (deterministic) conditional on $a$, $b$, and $\Om$, and determinism implies {\em statistical} (but not {\em functional}) independence, one could say that 
passive locality holds, but unrestricted active locality (equality in (\ref{xx_eqn}) does not. Alternatively, perhaps we should expand the definition 
of ``passive locality" to include statistical {\em and functional}
independence given $\Om$, for any functions appearing in the theory and affecting the outcome variables. So that is why
passive locality is violated, but not active locality.

\section{Computational Appendix\label{tech_chap}}

The random $v$'s were simulated using the method:

\bar
\no r_1 &\=& s\,\cos(\theta)\,\sin(\zeta);\\
\no r_2 &\=& s\,\sin(\theta)\,\cos(\phi)\,\sin(\zeta);\\
\no r_3 &\=& s\,\sin(\theta)\,\sin(\phi)\,\sin(\zeta);\\
\no r_4 &\=& s\,\cos(\zeta);
\ear

\ni where $\zeta$, $\theta$ were chosen uniformly in $[0,\pi]$ and $\phi$ uniformly in $[0,2\pi]$.
Then $v_1 = r_1\,\exp(\,i\,\omega_1\,)$, etc., where the $\omega_.$ were chosen uniformly in $[0,2\pi]$. All the angles were generated for each ``run" 
by the system-supplied Random Number Generator.

Concerning solving equations (\ref{macroeqns}) on the computer: define

\bar
\no q_1 &\=& X;\\
\no p_1 &\=& \dot{X};\\
\no q_2 &\=& Y/N;\\
\no p_2 &\=& \dot{Y}/N.\\
\no V(q_1,q_2) &\=& V_E(q_1) + \alpha\,N\,q_1\,q_2,\\
&&
\ear

\ni and a fake Hamiltonian:

\be
H_{\hbox{fake}} \= (1/2)\,p_1^2 \+ (1/2)\,p_2^2 + V(q_1,q_2) \+ f\,q_2.
\ee

\ni (Here $f$ stands for the random force, the expression with the $v$'s.)
Then (\ref{macroeqns}) are the Newtonian equations associated with this fake Hamiltonian. 
Although these equations constitute an ODE system, the usual methods such as Runge-Kutta (supplied in software packages) are not appropriate, since they do not conserve energy or the
volume-preserving property of Hamilton's equations. 
The proper technique for solving such equations is called a ``symplectic algorithm."
I used a method due to Ronald Ruth (1983), \cite{ruth}, which is also described (as of 2015) on the ``symplectic integrator" Wikipedia webpage. 
I tested the software on the Kepler problem as usual; provided the step size was small enough, the method preserved energy and angular momentum to five decimal places (and agreed with the exact solution). 
It also preserved energy
for the model simulated here. 

Parameters used were: $A = 2.0$; $R = 1.0$; $\alpha = .003$; the random forces were $\pm.003$; $N = 10$, and 10,000 time-steps per simulation. The formula for the quartic is:

\be
V_E(x) \= \left(\,\frac{A}{R^4}\,\right)\,x^4 \-  \left(\,\frac{2\,A}{R^2}\,\right)\,x^2 + A.
\ee

\vskip0.1in

The program was written in the C language in the style of the 1970s and run on a PC using the Linux operating system. If anyone
wishes to replicate the results here, that person should use a modern platform.
To make the figures, I used the shareware statistics package called R.

\end{document}